\documentclass[sigconf]{acmart}

\AtBeginDocument{%
  \providecommand\BibTeX{{%
    \normalfont B\kern-0.5em{\scshape i\kern-0.25em b}\kern-0.8em\TeX}}}

\setcopyright{acmlicensed}
\copyrightyear{2024}
\acmYear{2024}

\acmConference[CIKM '24] {Proceedings of the 33rd ACM International Conference on Information and Knowledge Management}{October 21--25, 2024}{Boise, ID, USA.}

\acmBooktitle{Proceedings of the 33rd ACM International Conference on Information and Knowledge Management (CIKM '24), October 21--25, 2024, Boise, ID, USA}

\acmISBN{979-8-4007-0436-9/24/10}
\acmDOI{10.1145/3627673.3680076}

\usepackage{balance}
\begin{document}

\title{Enhancing Playback Performance in Video Recommender Systems with an On-Device Gating and Ranking Framework}

\author{Yunfei Yang}
\affiliation{%
  \institution{Kuaishou Technology}
  \city{Beijing}
  \country{China}
}
\email{yfy@pku.edu.cn}

\author{Zhenghao Qi}

\affiliation{%
  \institution{Kuaishou Technology}
  \city{Beijing}
  \country{China}
}
\email{qizhenghao@kuaishou.com}

\author{Honghuan Wu}

\affiliation{%
  \institution{Kuaishou Technology}
  \city{Beijing}
  \country{China}
}
\email{wuhonghuan@kuaishou.com}

\author{Qi Song}

\affiliation{%
  \institution{Kuaishou Technology}
  \city{Beijing}
  \country{China}
}
\email{songqi@kuaishou.com}

\author{Tieyao Zhang}

\affiliation{%
  \institution{Kuaishou Technology}
  \city{Beijing}
  \country{China}
}
\email{zhangtieyao@kuaishou.com}

\author{Hao Li}

\affiliation{%
  \institution{Kuaishou Technology}
  \city{Beijing}
  \country{China}
}
\email{lihao@kuaishou.com}

\author{Yimin Tu}

\affiliation{%
  \institution{Kuaishou Technology}
  \city{Beijing}
  \country{China}
}
\email{yiming@kuaishou.com}

\author{Kaiqiao Zhan}

\affiliation{%
  \institution{Kuaishou Technology}
  \city{Beijing}
  \country{China}
}
\email{zhankaiqiao@kuaishou.com}

\author{Ben Wang}
\authornote{Corresponding author}

\affiliation{%
  \institution{Kuaishou Technology}
  \city{Beijing}
  \country{China}
}
\email{wangben@kuaishou.com}

\renewcommand{\shortauthors}{Yunfei Yang, et al.}

\begin{abstract}
Video recommender systems (RSs) have gained increasing attention in recent years. Existing mainstream RSs focus on optimizing the matching function between users and items. However, we noticed that users frequently encounter playback issues such as slow loading or stuttering while browsing the videos, especially in weak network conditions, which will lead to a subpar browsing experience, and may cause users to leave, even when the video content and recommendations are superior. {\bfseries It is quite a serious issue, yet easily overlooked}.

To tackle this issue, we propose an on-device \textbf{G}ating and \textbf{R}anking \textbf{F}ramework (\textbf{GRF}) that cooperates with server-side RS. Specifically, we utilize a gate model to identify videos that may have playback issues in real-time, and then we employ a ranking model to select the optimal result from a locally-cached pool to replace the stuttering videos. Our solution has been fully deployed on Kwai, a large-scale short video platform with hundreds of millions of users globally. Moreover, it significantly enhances video playback performance and improves overall user experience and retention rates.

\end{abstract} 


\maketitle

\section{Introduction}

The popularity of short video platforms, such as TikTok, YouTube Shorts and Kuaishou, has led to a surge in video recommendation technology over the past few years. While significant progress has been made in this field, most of them focused on matching and sorting algorithms, including both cloud-based methods \cite{lin2022feature, zhan2022deconfounding, zhou2018deep} and on-device solutions \cite{gong2020edgerec, gong2022real,feng2021grn} that utilize real-time signals for addressing the delays caused by server-side paging mechanisms.

\begin{figure}[h]
  \centering
  \includegraphics[width=\linewidth]{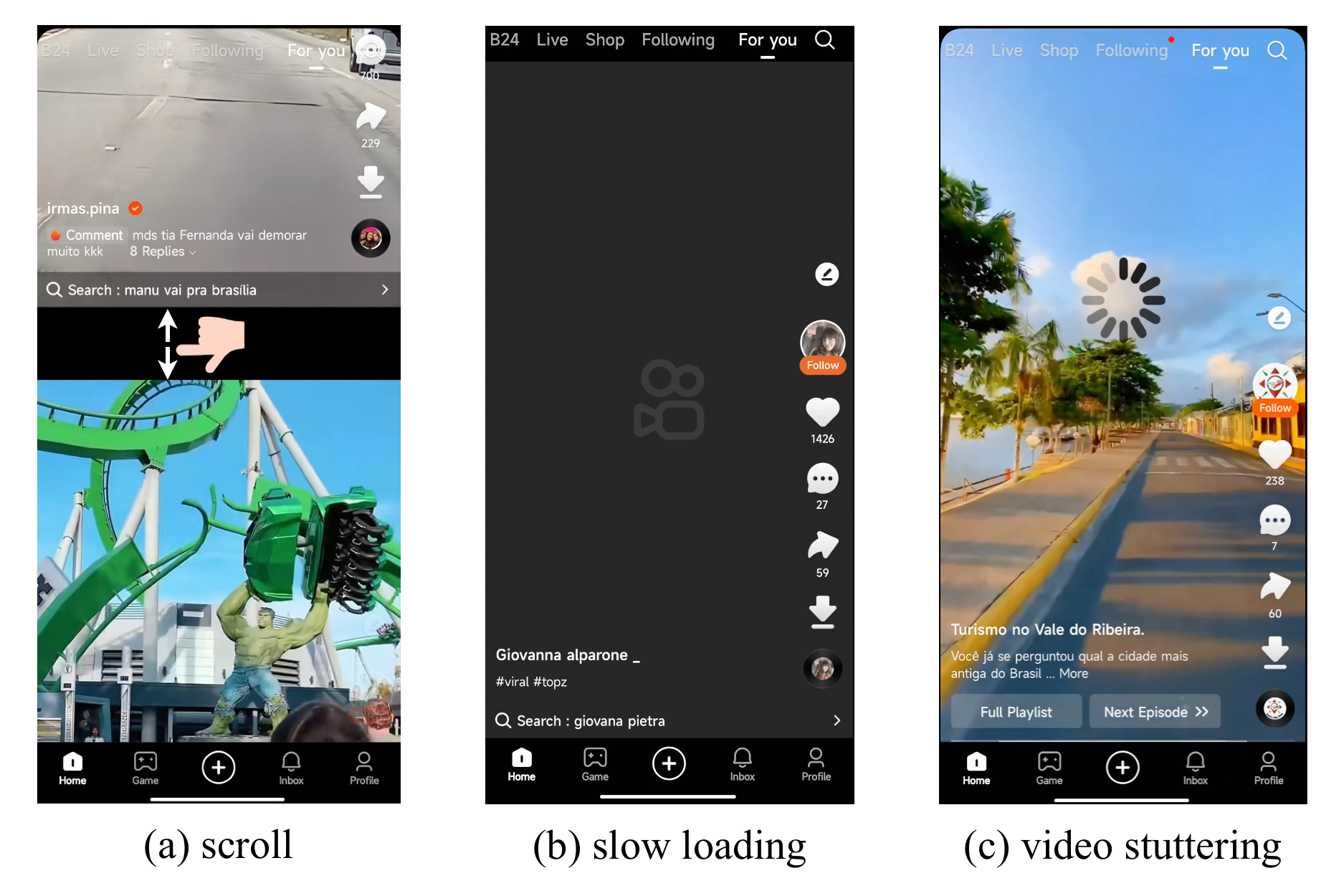}
  \caption{An example interface of short video platform. (a) Users browse videos by scrolling up and down. (b) and (c) illustrate several playback issues while browsing the videos.}
  \label{fig:fig1}
\end{figure}

In real-world scenarios, users often face playback issues such as slow loading or stuttering (referred to as "choppy") as illustrated in Figure \ref{fig:fig1}, This is especially prevalent in underdeveloped regions or areas with unstable network conditions. For instance, \textbf{71\% } of global users on Kwai have encountered subpar network conditions, leading to an average of 6-7 instances of choppy playback per day. This problem can significantly diminish the user experience and may cause users to leave the platform. Improving prediction accuracy, as the aforementioned algorithms aim to do, cannot completely mitigate this issue.

To evaluate the impact of playback performance on user experience, we select several representative playback issues, such as failed loading, slowing loading of the first screen, and video stuttering as choppy samples. Figure \ref{fig:vv} shows the relationship between average session watch time and the proportion of choppy videos within a session (choppy rate). We can conclude that the average session watch time significantly decreases as the choppy rate increases. Specifically when the choppy rate reaches 10\%, the average session watch time drops by 56.1\%. Therefore, it is evident that playback performance has a substantial impact on user experience, and addressing playback issues holds considerable potential for improving user engagement.

Previous research has also noticed the issue of network influence. For example, \cite{lang2023beyond} attempts to jointly consider video attractiveness and network states to improve the overall revenue. However, most of these efforts were server-side, lacking the ability to adjust content based on real-time signals from the client.  choppy playback is a multifaceted issue caused by complex factors. Such as $(1)$ {\bfseries Network Environment}: Real-time traffic fluctuations due to volatility in network performance can cause playback lag behind video loading progress. $(2)$ {\bfseries Device Status}: Resource contention and execution of time-consuming tasks caused by running multiple apps on the client can also cause stuttering. and $(3)$ {\bfseries Content Quality}: Some high-bitrate videos require specific rendering and loading processes.

Enhancing playback performance is a real-time client scenario, where device and network status are subject to dynamic changes. Thus, the best solution is to address these issues on the client side. Numerous factors can contribute to playback performance, making it difficult to assess through simple strategies. Misjudging these factors can lead to replacing a high-quality video, thereby diminishing the user experience.

In this paper, we propose a novel on-device gating and ranking framework for enhancing video playback performance. In particular, to screen out the choppy instances accurately, we deploy a gate model on device to evaluate the playback status of upcoming video in real-time. Furthermore, we incorporate a ranking model to select the optimum result from a locally-cached pool for replacement of the choppy ones. To strengthen the attractiveness of cached videos, we dynamically maintain a personalized pool, where the videos are obtained from server-side recommender system via a cache update mechanism.

Our contributions can be summarized as follows:

\begin{figure}[h]
  \centering
  \includegraphics[width=\linewidth]{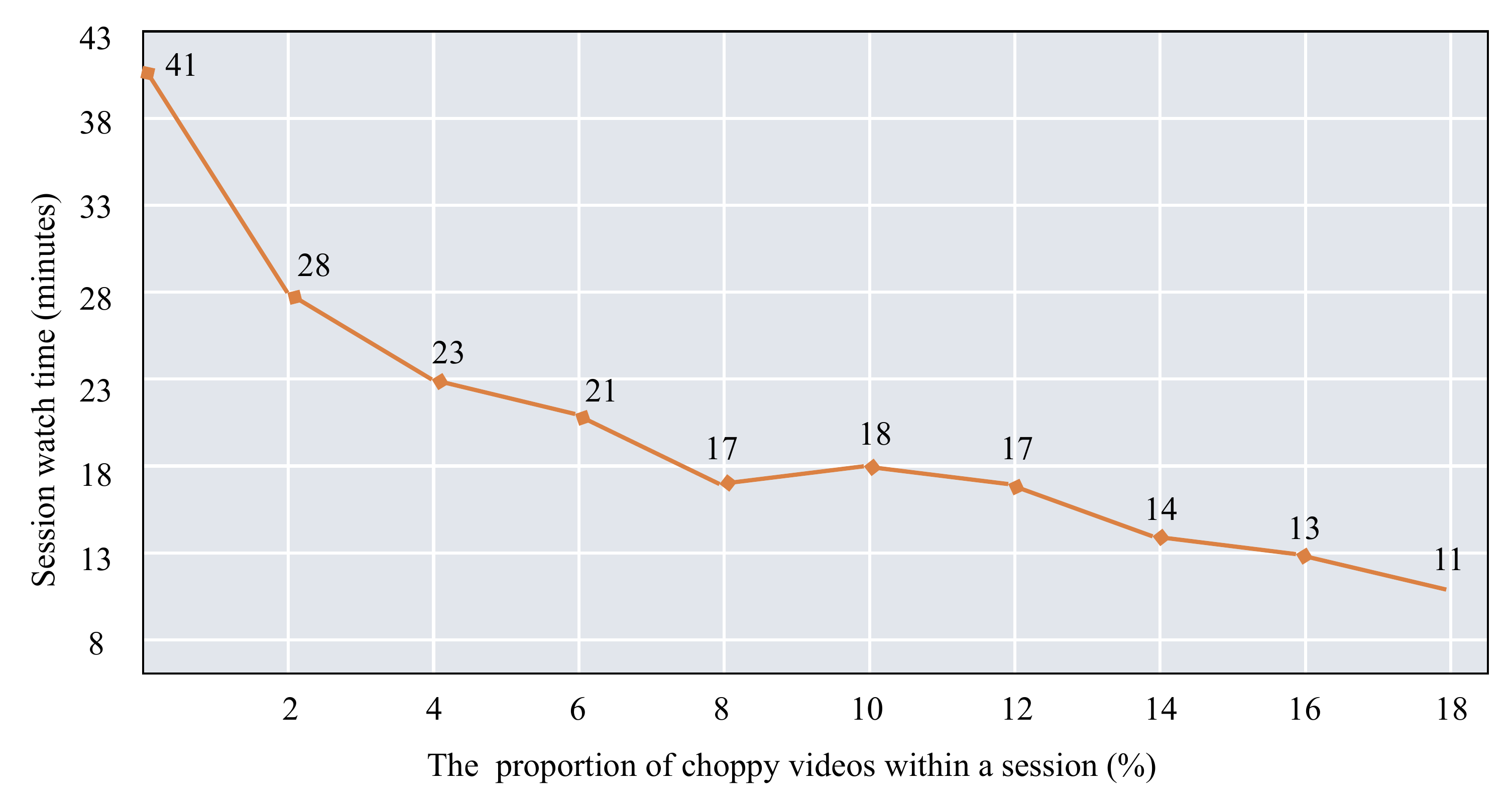}
  \caption{The relation between average session watch time and the proportion of choppy videos in a session.}
  \label{fig:vv}
\end{figure}

\begin{itemize}
\item We highlight the choppy playback, a real issue experienced by the client, and demonstrate its substantial impact on video recommender systems. This indicates a prospective direction for improving user engagement. 
\item We propose a simple yet effective on-device gating and ranking framework, called GRF, for enhancing playback performance in video recommender systems. To the best of our knowledge, we are the first to provide an on-device RS in a network-free environment, and it is easy to adapt our design to other scenarios.
\item We conducted extensive offline experiments and online A/B testing in a real-world industrial scenario, which further demonstrate the effectiveness of our framework. 
\end{itemize}

\section{RELATED WORK}

\subsection{On-device Recommender Systems}


Most industry-scale recommender systems (RSs) were deployed on cloud server with a multi-stage paradigm. Typically, RSs compute the top-N items and transmitted an ordered list to client from each request, thus it cannot adjust the results between two adjacent requests. To fully utilize client-side real-time signals, some RSs have been successfully deployed on-device. For example, EdgeRec {\cite{gong2020edgerec}} propose a context-aware reranking with behavior attention network to model interactions between candidate items and real-time user behavior. \cite{gong2022real} design a re-ranking solution that fits on mobile devices for short-video recommendation. These methods emphasize tackling system latency caused by server-side paging mechanisms and capturing real-time user preference. However, We highlight the importance of playback performance and address it on-device in a real-world video recommender system.

\begin{figure*}[!htb]
  \centering
  \includegraphics[width=\textwidth]{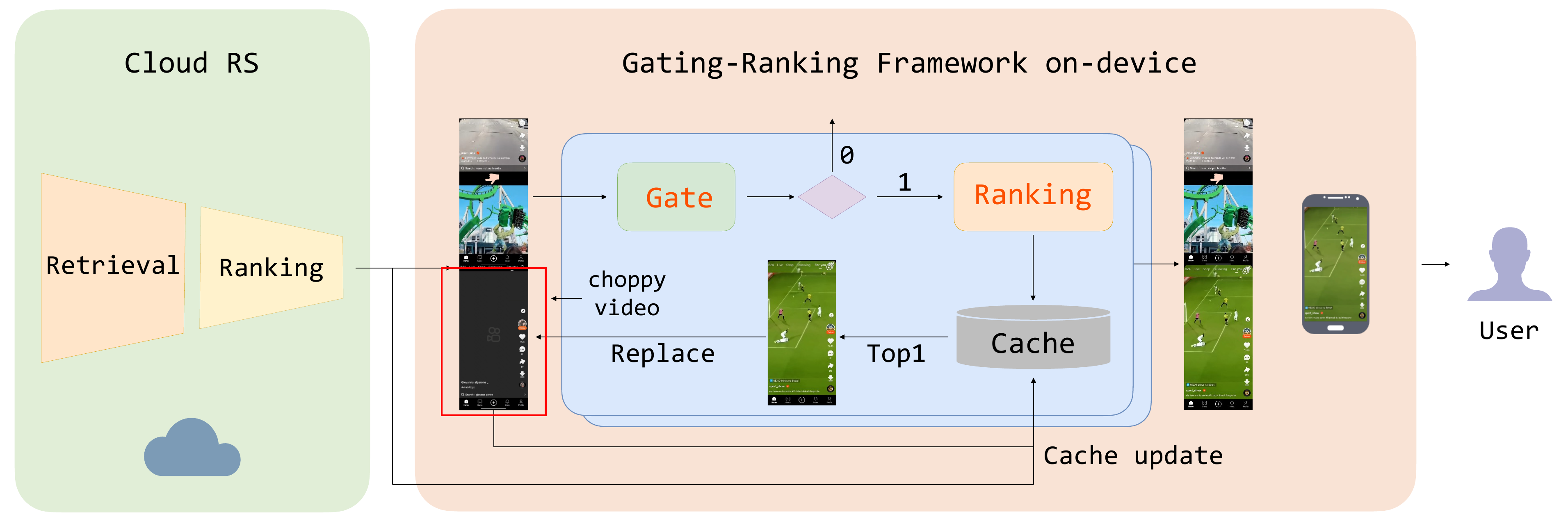}
  \caption{The proposed Gating and Ranking framework (GRF) involves two stages: Gating, which screens out choppy video (surround with a red rectangle), and Ranking, which selects the best local cache video for the replacement.}
  \label{fig:system}
\end{figure*}

\subsection{Playback Performance Optimization}
Due to the volatility of network conditions and the potential variability in video sizes, users frequently encounter choppy playback while browsing the videos. To address this, previous works {\cite{akhtar2018oboe, feng2019vabis, hodzic2022realistic}} employ a video bit rate adjustment algorithm, which adaptively adjusts the video bit rate during playing to improve playback performance. Through effective, the main limitations lie in (1) unable to handle the failed loading or slow loading of the first screen, (2) some user may be sensitive to video quality adjustment. \cite{lang2023beyond} propose a recommendation framework which jointly considers network status and video attractiveness. Since the solution works on the cloud server, it lacks the ability to adjust recommendation contents based on real-time playback performance.

\section{SYSTEM}


In this section, we present an overview of the proposed system, As illustrated in Figure \ref{fig:system}, GRF works on the device and seeks to cooperate with RS on cloud, the system involves two stages:  $(1)$ Gating and $(2)$ Ranking. The gating operates when a user scrolls down on the screen and makes an inference for the upcoming video. Once the video is predicted to be a choppy sample,  the ranking stage will automatically activate and select an optimal result from a personalized locally-cached pool to replace the choppy ones with real-time features.



\subsection{Problem Definition}

%

We formulate the GRF as a two-stage problem, whose input is the ranking list requested from the server side, represented as $V = \left\{v_1, v_2, ...,v_n\right\}$. Let $C = \left\{c_1, c_2, ...,c_{|C|}\right\}$ denote the videos in the locally-cached pool. The problem can be defined as follows: given a video sequence $V$, user $u$ and corresponding candidates set $C$. We first train a binary classifier to determine the output value $y_i \in \left\{ 0, 1\right\}$ for each upcoming video $v_i \in V$ , where $y_i$ represents the prediction status of $v_i$. Once $y_i=1$, the ranking model is activated to choose an optimal video $c_j \in C$ via a multi-task learning paradigm. After that, the video $c_j$ will be displayed to the user, while the video $v_i$ will queued for caching and stored in the locally-cached pool. The overall notations are summarized in Table \ref{tab:not}.

\begin{figure*}[!htb]
  \centering
  \includegraphics[width=\textwidth]{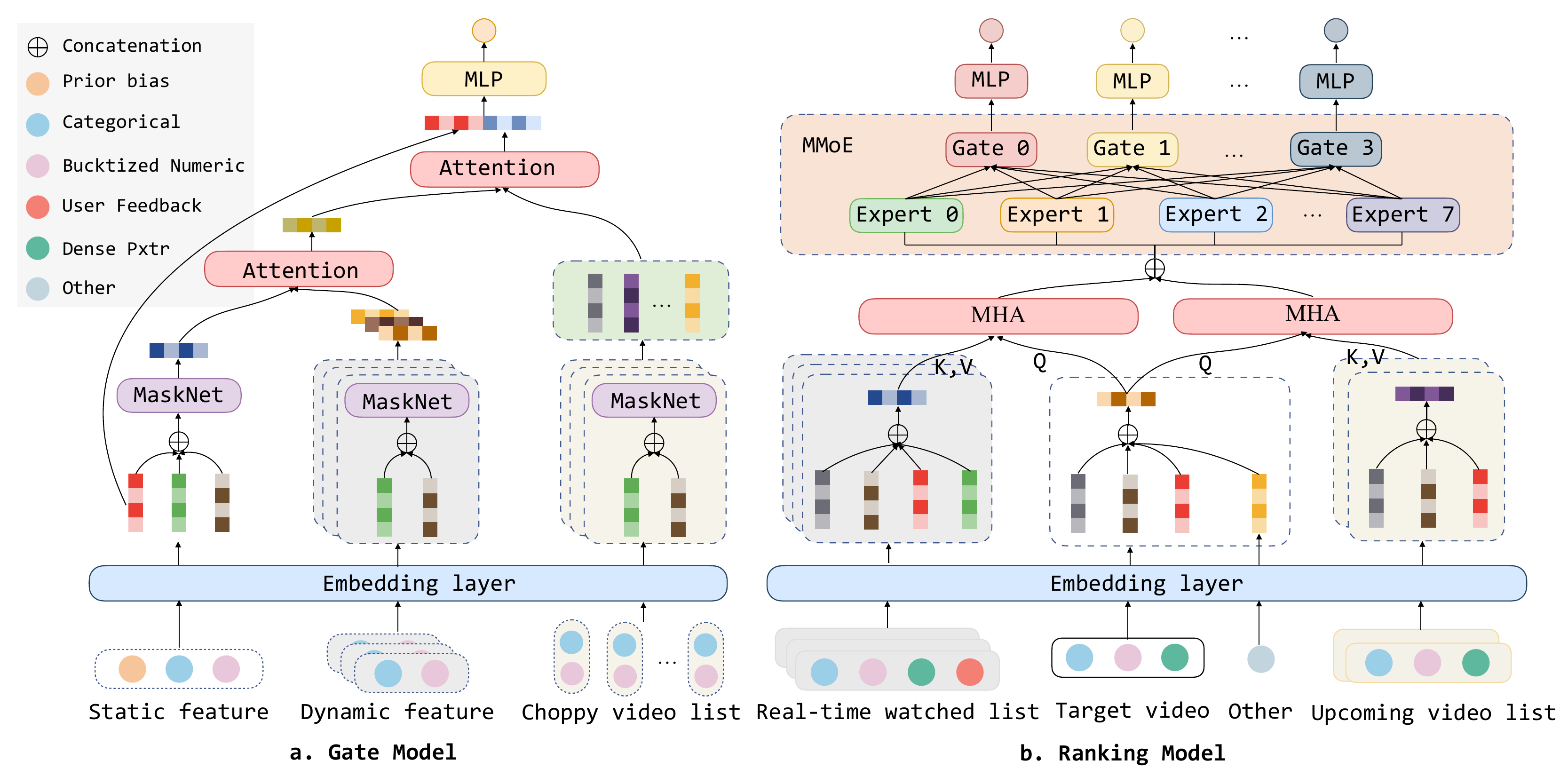}
  \caption{The architecture of the on-device Gate model (a) and Ranking model (b).}
  \label{fig:fig3}
\end{figure*}

\subsection{Stage \uppercase\expandafter{\romannumeral1} : Gating}

The goal of the gate model is to screen out the choppy videos. Based on the video playback status, we define cases such as video stuttering, slow loading of the first screen, and failed loading as positive (or choppy) samples, while seamless playback is considered a negative sample. The model structure is presented in Figure \ref{fig:fig3} (a),

{\bfseries Features.} We elaborately pick out and extract features to keep the model lightweight yet effective. Input Features consist of several categories: $(a)$ Static feature, such as video duration, device score, prior bias, etc. Prior bias consists of several categorical features classified with statistical data manually. For example, the video's size may divided into 3 levels: small, middle, and large, which can mapped to 0, 1, 2.   $(b)$ Dynamic feature (such as network speed, video cached ratio, etc.). These features will collected during a real-time watched list and constructed as a sequence. and $(c)$ Choppy video list $H_c$, the choppy video sequence of the user's watching history. To reduce the dimension of feature space, we bucketize the numerical features to equal distance bins. Following \cite{gong2022real}, we then embed the discrete features to latent space using AutoDis \cite{guo2021embedding} technique.

\begin{table}[htbp]
  \caption{The main notations and features of our work.}
  \label{tab:not}
  \begin{tabular}{cc}
    \toprule
    Notation & Description \\
    \midrule
    $u$ & a user \\
    $v$ & a video \\
    $V$ & Ordered video list from cloud RS \\
    $C$ & The set of videos in locally-cached pool. \\
    $R$ & latest video list in user's watched history. \\
    $H_c$ & Choppy video list in user's watched history. \\
    $s_{net}$ & current network speed. \\
    $V_{up}$ & Upcoming video list. \\
    $v_{dur}$ & The cached duration of a video.\\
    $v_{rat}$ & The cached ratio of a video.\\
    $pxtr$ & The predicted value of server-side ranking model.\\
    $fpr$ & the predicted rate of completely watching video.\\
    $evr$ & the predicted rate of effectively watching video.\\
    $lvr$ & the predicted rate of long watching video.\\
    $svr$ & the predicted rate of watching less then 3 seconds.\\
  \bottomrule
\end{tabular}
\end{table}

{\bfseries MaskNet \& Target attention.} We adopt a parallel MaskNet layer \cite{wang2021masknet} to capture complex high-order features effectively. This layer takes the feature embedding concatenation as input and exploits multiple MaskBlock to incorporate different kinds of features and their interactions. Then, we employ a hierarchical target attention \cite{zhou2018deep} to attend most important information in feature sequence. The first attention layer is used to capture feature evolution with time series, while the second layer is utilized to highlight the informative features in history choppy video list. 

{\bfseries Output Layer.} Let $h$ be the final hidden representation of the last target attention layer, we concatenate $h$ with the representation of the prior bias feature. Then, we employ a linear layer followed by a sigmoid activation function to predict whether the target video is choppy.




\subsection{Stage \uppercase\expandafter{\romannumeral2} : Ranking}

The structure of ranking model is presented in Figure \ref{fig:fig3} (b). Different from gating stage, the goal of the Ranking model is to predict multiple xtrs and can be formulated as a multi-task learning problem. \cite{caruana1997multitask}. Let $v_i \in V$ be a choppy video identified by discriminator and $v_{i-1}$ be impression video at present, the input features of the model are as follows:

{\bfseries Real-time watched list $R$.} The latest videos in user's watched history,  The information of these videos contains categorical features, dense pxtrs from server-side model, bucketized numeric and user feedback, etc.

{\bfseries Target video.} Target video and its information are stored in locally-cached pool. 

{\bfseries Upcoming video list. $V_{up}$} Since ranking model aims to select a video to replace $v_i$, the remaining videos $ \left\{v_{i+1}, v_{i+2}, ...,v_n\right\}$ of original ranked list are also be feed into model as a contextual feature to capture mutual influence between different items.

{\bfseries Multi-head Target Attention.} We incorporate a multi-head attention mechanism (MHA) \cite{vaswani2017attention} to capture the interactions between target video and real-time watched sequence. In this context, we have replaced self-attention with target attention proposed in \cite{zhou2018deep}.
This layer takes target video representation $h \in R^{1 \times d_h} $, real-time watched sequence representation $s \in R^{l \times d_s} $ as input and output a vector $o \in R^{1 \times d_o}$ to represent user's interest. Specifically, the multi-head attention layer is computed as:

\begin{displaymath}
   MultiHead(Q, K, V) = Concat(h_1, ..., h_{n})W^O \nonumber 
\end{displaymath}

where $h_i \in R^{1 \times d_v}$ is the output of a single head, $W^O \in R^{n·d_v \times d_o}$ is the projection parameter matrix.

We compute each $h_i$ with a scaled dot product attention, the equation as:

\begin{displaymath}
   Attention(Q, K, V) = softmax( \frac{QK^T}{\sqrt{d_k}})V
\end{displaymath}

where $Q$ is projected from target video representation $h$, $K$ and $V$ is projected from sequence representation $s$ respectively.

Then we compute the contextual representations $o$ as $o = ew_o^T$, where $e \in R^{1 \times d_v}$ is the concatenation of multi-head representation, $w_o \in R^{d_o \times d_v}$ is the projection matrix of output. 


{\bfseries MMoE.} We utilize a multi-gate mixture-of-experts \cite{ma2018modeling} layer to incorporate various user feedbacks and model the correlation between different objectives. Each expert pays attention to specific kind of important features, and the fusing information in mixture-of-experts was combined through different gating networks for each task.  Finally, the prediction of task $i$ is:

\begin{displaymath}
    pxtr_i = sigmoid_i(MLP_i(x_{gate_i}))
\end{displaymath}

where pxtr donates the prediction of a certain objective, such as effective\_view, long\_view, etc. We calculate the final score of each video by a weighted sum of multiple pxtrs, and choose the video with the highest score for display.

\subsection{Cache update Mechanism}
Cache updated in the following situations: $(1)$ Choppy video is automatically added to the cache pool after being replaced. $(2)$ If the video's caching time exceeds 2 weeks, it will be removed from the cache pool. $(3)$ Each user has a local cache with the capacity of $C_{u}$ where $C_{u}$ is the max daily cache consumption in recent 7-days . If the current cached video is less than $0.75*C_{u}$, the client will ask for cache replenishment from the cloud. In our system, the average number of cached videos per user is 24.


\subsection{Recommendation in offline mode.}


It is worth mentioning that our system can also work effectively even without an internet connection, which we refer to as offline mode. In this scenario, since the client is unable to interact with the cloud server, we only perform the ranking stage for videos in the locally-cached pool. Offline mode continues until network connectivity is restored or the user has viewed all cached videos.

\subsection{System Implementation}

\begin{figure}[h]
  \centering
  \includegraphics[width=\linewidth]{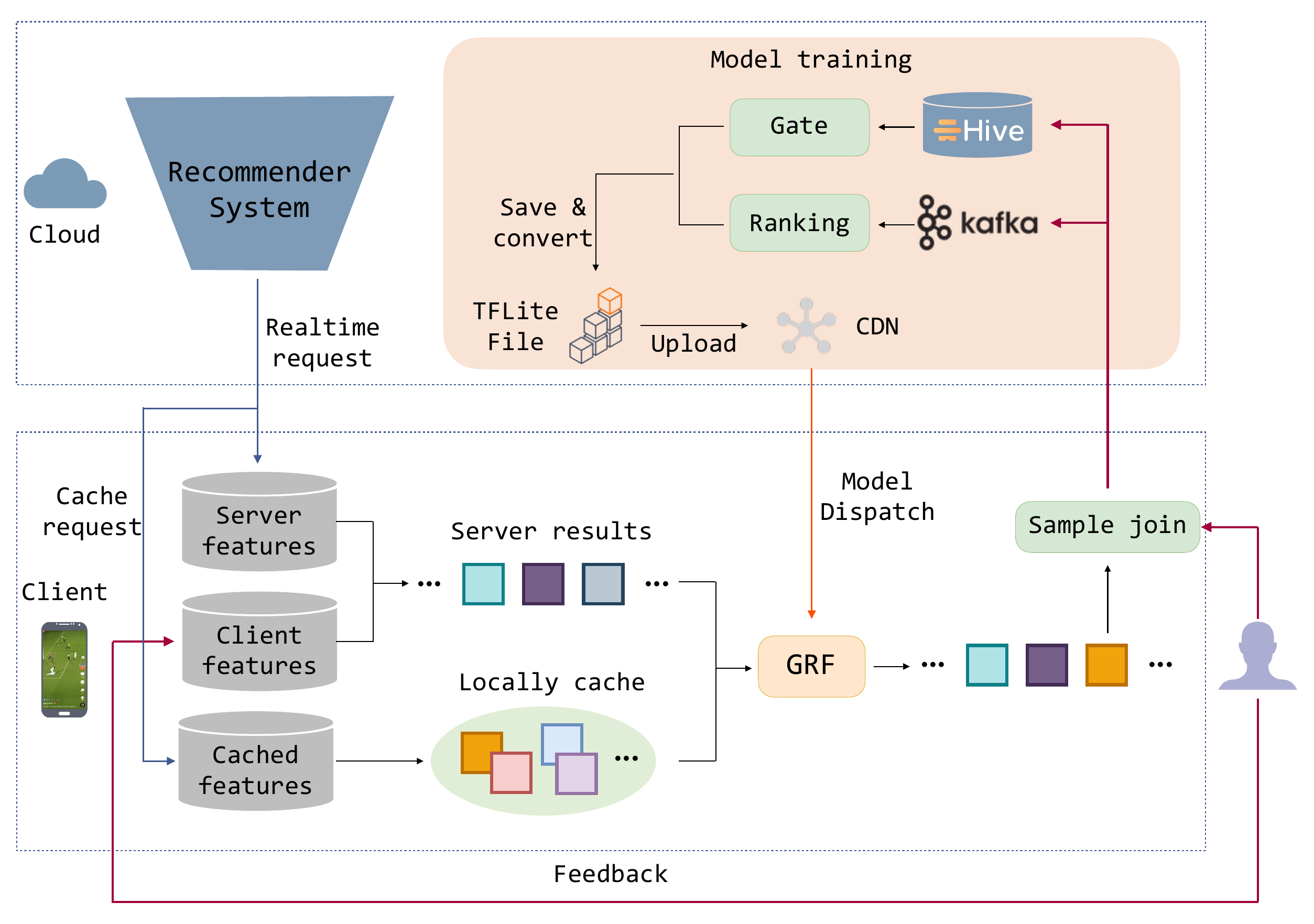}
  \caption{GRF System Implementation.}
  \label{fig:imp}
\end{figure}

We have deployed our proposed GRF in a real-world industry-scale video recommendation system. The structure of the online system is depicted in Figure \ref{fig:imp}, and we will introduce the system implementation in three parts: model training, deployment and inference, and system efficiency.

{\bfseries {\itshape Model Training} } When users browse the videos, we combine features (such as network speed, video cached duration, bit rate, etc.) with labels (such as whether the video is choppy and the user engagement metrics like watch time, likes, follows, comments, etc.) to create a complete sample. These samples will be stored in a Hive database \cite{thusoo2009hive}. We collect an offline dataset from the Hive table to train the gate model. Additionally, we use Kafka streams  \cite{narkhede2017kafka} to train the ranking model via an online learning paradigm, and we save the checkpoint of ranking model every four hours.


{\bfseries {\itshape Deployment and Inference} } The deployment of a checkpoint to the mobile device involves three steps: \textbf{step $1$}: Convert the TensorFlow checkpoint model into TFLite \cite{abadi2016tensorflow} format. \textbf{step $2$}: Upload the compressed TFLite format file to CDN. \textbf{step $3$}: The client downloads the model from CDN. We update the ranking model periodically with the latest checkpoints to maintain its real-time performance. For inference, the client collects features from both server-side and client-side, feeds the features into TFLite model, and makes recommendations based on the model's output.



\begin{table}[htbp]
  \caption{the efficiency metrics of GRF.}
  \label{tab:eff}
  \begin{tabular}{cccc}
    \toprule
    model & parameter & storage (TFLite) & inference time\\
    \midrule
    Gate & 654757 & 2.727M  & 43ms\\
    Ranking & 1310860 & 5.318M & 69.7ms\\
  \bottomrule
\end{tabular}
\end{table}

{\bfseries {\itshape System Efficiency} } Given the limited computing power, energy, and storage on mobile devices, resource consumption is a key metric for evaluating the system. the computing efficiency metrics of our system are listed in Table \ref{tab:eff}. It is evident that both models are sufficiently lightweight in terms of storage and inference time, allowing them to run smoothly without a significant performance deterioration on the user's mobile device.

\section{EXPERIMENTS}

In this section, we conduct offline and online experiments with real-world video RS dataset to demonstrate the effectiveness of our proposed framework.

\subsection{Offline Results of Gate model}

{\bfseries {\itshape Experiment Setup.} } The goal of the gate model is to screen out choppy videos, which can be formulated as a binary classification task. We use an exponential decay learning rate schedule which keeps the value for the first 50 steps, the initial learning rate is 7e-5, and the decay rate is 0.98. We apply a class weight of 10:1 for positive and negative samples respectively to alleviate the imbalanced issue. the batch size is 2048 and the number of maskBlocks is 3.

{\bfseries {\itshape Dataset} }We collected data from online recommendation logs in Kwai from 2023-09-11 to 2023-09-17. There are a total of 12368053 impressions consisting of 386501 choppy samples. We randomly split data into training, validation, and testing three parts with the ratio of 8:1:1, respectively.

\begin{table}[htbp]
  \caption{offline results of different models.}
  \label{tab:dis}
  \begin{tabular}{ccc}
    \toprule
    model & AUC &  recall@p0.7\\
    \midrule
    Xgboost & 0.847 & 0.624 \\
    SimpleDNN & 0.887 & 0.651 \\
    TabTransformer & 0.919 & 0.675 \\
    Ours & \textbf{0.957} & \textbf{0.692} \\
  \bottomrule
\end{tabular}
\end{table}

{\bfseries {\itshape Evaluation Metrics and Baselines.} } We use AUC and recall at precision 0.7 as evaluation metric. We compare our model with three representative baselines:  $(1)$ \textbf{Xgboost}.  A widely used gradient boosting decision tree algorithm \cite{chen2016xgboost}. $(2)$ \textbf{SimpleDNN}. A DNN model with embedding strategy in our work. and $(3)$ \textbf{TabTransformer}. \cite{huang2020tabtransformer} Which is a novel transformer architecture for modeling tabular data. We implement it with pytorch\_tabular toolkits\cite{joseph2021pytorch}. 

We evaluation recall@p0.7 as this threshold achieves a good balance between cache costs and overall benefits in online experiments. Table \ref{tab:dis} illustrates that our gate model significantly outperforms other baselines including both machine learning and deep learning algorithms, which shows the effectiveness of our methods.

\subsection{Offline Results of Ranking model}

{\bfseries {\itshape Experiment Setup.} } Our ranking model aims to predict the ratio of four objectives: effective\_view (evr), long\_view (lvr), short\_view (svr) and finish\_play (fpr), each indicating varying degrees of user engagement during watching. We use 4 attention heads in both real-time and upcoming video list MHA layer, and the dimension of each head is 64. The MMoE module consists of 8 experts and 4 gates. We use Adam \cite{kingma2014adam} to optimize the model parameters, the batch size is 4096 and the learning rate is 4e-5.

{\bfseries {\itshape Evaluation Metrics and Baselines.} } Following \cite{gong2022real}, we use AUC as main metric. When ranking cached videos, we have two options: 1. use the pxtr from the server-side model, or 2. use our ranking model. Therefore, we take the server-side model as baseline. Since the model architecture is not the main contribution of this paper, and some state-of-the-art cloud-based models are difficult to deploy on devices due to resource constraints, we did not conduct additional experimental comparisons with more SOTA models.




\begin{table}[htbp]
  \caption{offline AUC results. The bold number indicates statistical significance improvement compared to baseline, measured by t-test with p-value < 0.05. }
  \label{tab:rank}
  \begin{tabular}{ccccc}
    \toprule
    model & evr & lvr & svr & fpr \\
    \midrule
    Server-side Model& 0.8135 & 0.8340 & 0.8087 & 0.8705 \\
    Our Ranking Model & \textbf{0.8170} & \textbf{0.8372} & \textbf{0.8161} & \textbf{0.8756} \\
  \bottomrule
\end{tabular}
\end{table}

As shown in table \ref{tab:rank}, our lightweight ranking model achieves superior performance compared with server-side large model, showing the ability of our method to select optimal videos. An intuitive explanation is that we jointly leverage the output of server-side model and real-time client features. 



\subsection{Online A/B Testing}

We conducted online A/B Testing in production environment from 2023-08 to 2023-10. We divide experiments into three settings. \textbf{1. base}: a blank control group without GRF system, \textbf{2. gate}: a group with gate model, and rank the cached video with server-side pxtr. and  \textbf{3. full}: a group with full proposed methods. We attribute 10\% traffic for each experiment group. 

\begin{figure}[h]
  \centering
  \includegraphics[width=\linewidth]{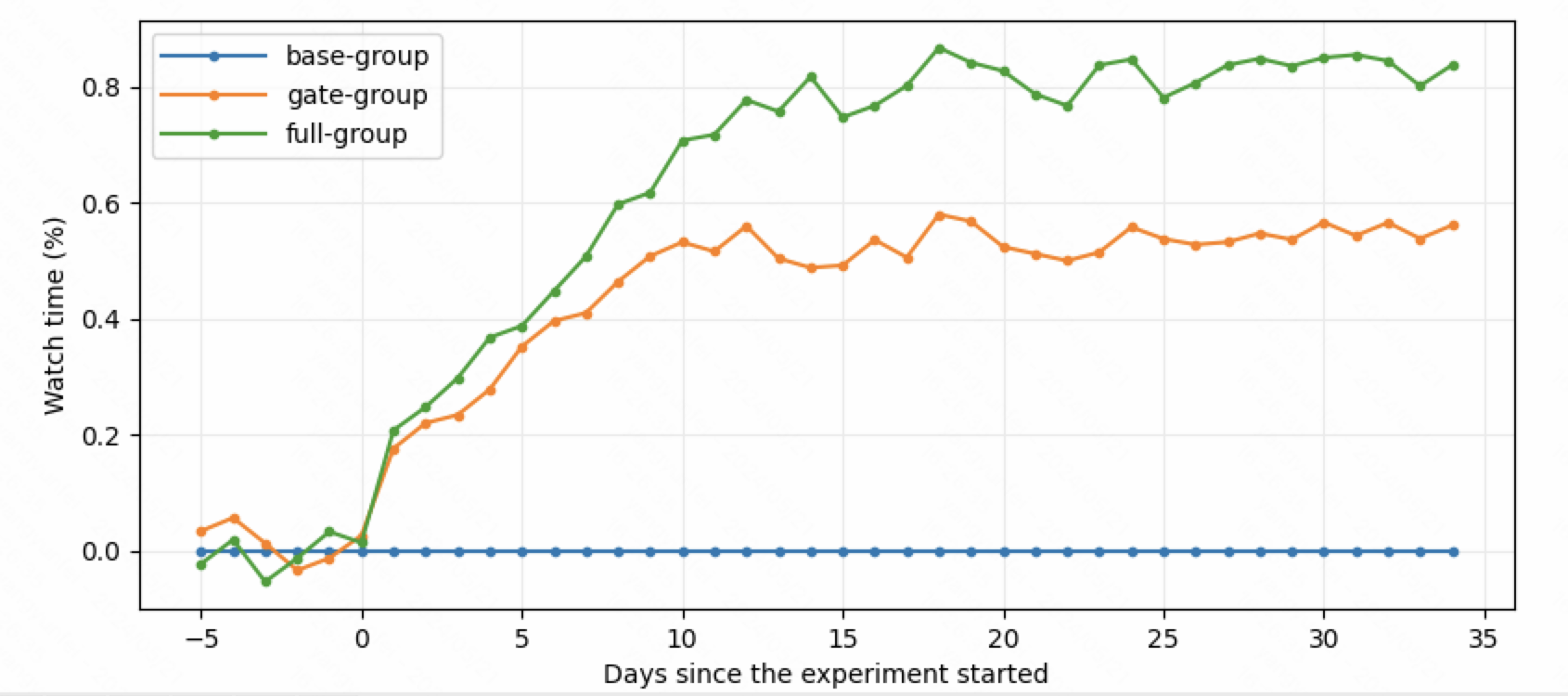}
  \caption{The online performance of different experiment group.}
  \label{fig:time}
\end{figure}

Figure \ref{fig:time} plots the online performance of different experiment group, where the x axis stands for the number of days since experiment started, and y axis denotes the percentage improvement of overall watch time. After over a month's rigorous testing, we report the results in table \ref{tab:abtest}, where the data is taken from the average values of the last seven days in online experiment. 



\begin{table}[htbp]
  \caption{Results of online A/B testing, measured by overall watch time (WT), watch time in poor network conditions (WTP), Day 1 Retention Rates (D1RR), Day 7 Retention Rates (D7RR), and Day 30 Retention Rates (D30RR). Here bold text indicates the best results, while * donates statistical significance improvement, measured by t-test with p-value < 0.05.}
  \label{tab:abtest}
  \begin{tabular}{cccccc}
    \toprule
     & WT & WTP & D1RR & D7RR & D30RR  \\
    \midrule
    base& +0.00\% & +0.00\% & +0.00\% &+0.00\% & +0.00\%\\
    \bottomrule
    gate& +0.551\%* & +31.57\%* & +0.134\%* &+0.122\% & +0.141\%*\\
    Full & \textbf{+0.839\%}* & \textbf{+44.45\%}* & \textbf{+0.141\%}* &\textbf{+0.136\%}* & \textbf{+0.157\%}* \\
  \bottomrule
\end{tabular}
\end{table}


Results in table \ref{tab:abtest} manifest that both Gate model and full GRF system achieves a considerable gains. Compare with base group, our method improved the overall watch time by 0.839\%, and watch time in poor network conditions (slow internet speed or weak network signal) by a large margin \textbf{44.45\%}, which further demonstrates the effectiveness of our methods in real-world scenarios. Our solution has been deployed online from then on.

\subsection{Case Study}

\begin{figure}[h]
  \centering
  \includegraphics[width=\linewidth]{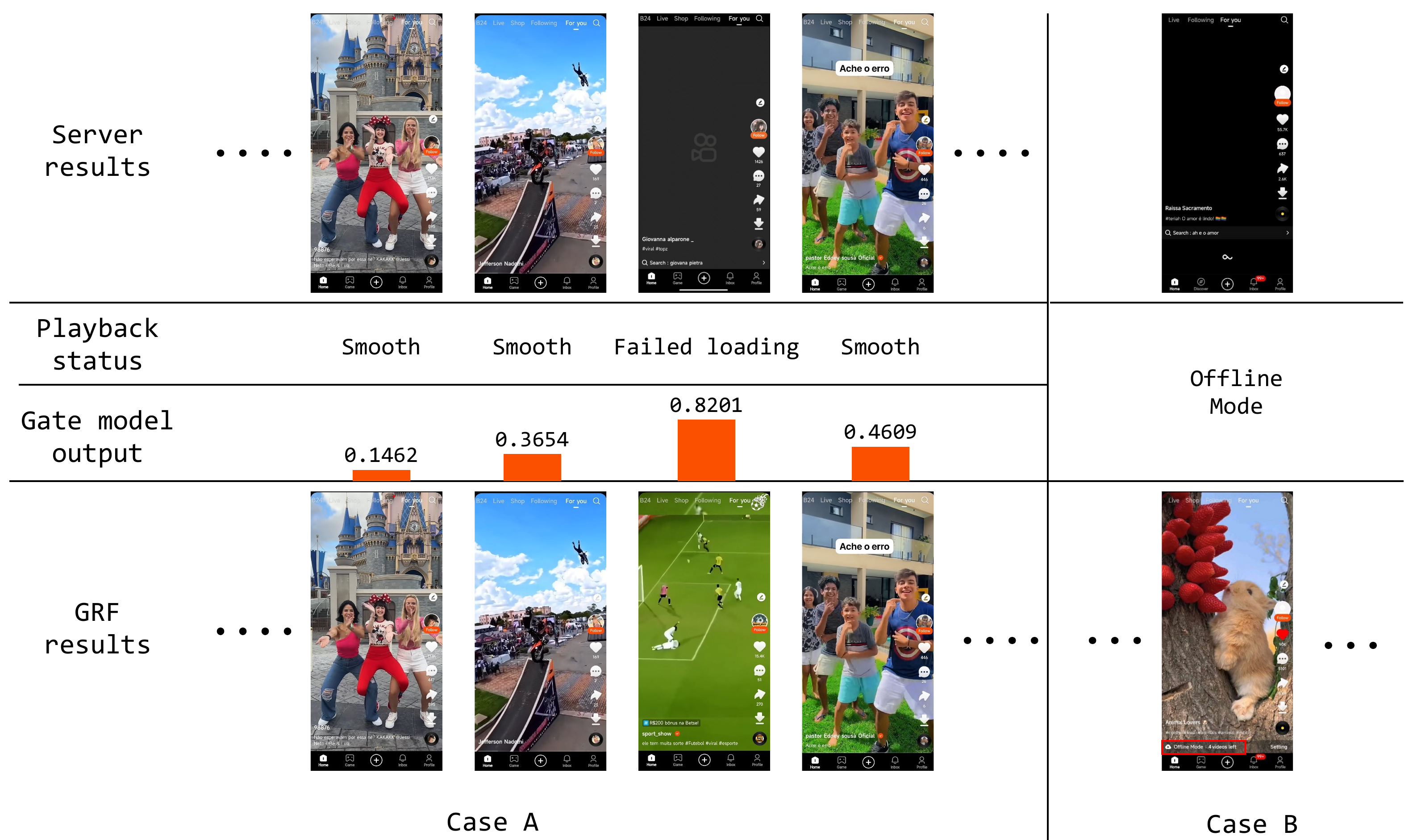}
  \caption{Two real-world cases to illustrate the impact of
our proposed gating and ranking framework.}
  \label{fig:case}
\end{figure}

We present two real-world cases to illustrate the impact of our proposed gating and ranking framework, as shown in Figure \ref{fig:case}. The top row displays the server-side recommendation results, the bottom row indicates the results of our GRF system, and the remaining rows show the video playback status and predicted value of the gate model. Once the GRF is implemented, the actual playback status of the replaced video is no longer known. In these two cases, we only obtain the GRF model's predicted results without making any changes to the recommendation contents. 

For case A, the third video encounters a failed loading playback issue, the predicted choppy rate of this video from the gate model is 0.8201, which is greater than the threshold of our online settings (0.75), thus our GRF system will replace it with a locally-cached video and dissolve this playback issue case. Case B shows another representative scenario, The user lost the network connection, making it impossible to interact with the cloud server and watch videos seamlessly.
However, our GRF system will switch to offline mode and make recommendations from a locally-cached pool, providing users with the ability to watch videos temporarily.

\section{CONCLUSION}

In this paper, We highlight the choppy playback, a real issue experienced by the client and show its substantial impact on user experience in video recommender systems. Furthermore, we propose a gating and ranking recommendation framework, called GRF, for enhancing playback performance, which makes the first attempt to provide an on-device RS that tackles the issue of choppy playback. Specifically, GRF involves two stages: the gating stage, which aims to identify videos that may have playback issues, and the ranking stage, which seeks to select the optimal result from a locally-cached pool to replace the choppy videos. We conduct extensive offline experiments and online A/B testing in a real-world industrial scenario to demonstrate the effectiveness of our methods. Currently, our solution has been fully deployed on mobile Kwai and significantly improved overall user watch time and retention rates.


We believe that tackling the issues in real-world recommender systems, such as choppy video playback, is a promising direction in the future. We will explore the collaboration and consistency between the recommendation models on the cloud and the device for further work.

\bibliographystyle{ACM-Reference-Format}
\balance
\bibliography{sample-base}

\appendix

\end{document}